\documentclass[aps,pre,reprint,superscriptaddress]{revtex4-2}
\usepackage{eurosym}
\usepackage{graphicx}
\usepackage{amsmath}
\usepackage{latexsym}
\usepackage{amsthm, amscd, amsfonts, amssymb}
\usepackage{multirow}
\usepackage{makecell}
\usepackage{color}

\begin{document}

\title{Ultra-high-amplitude Peregrine solitons induced by helicoidal
spin-orbit coupling}
\author{Cui-Cui Ding}
\affiliation{Research Group of Nonlinear Optical Science and Technology, Research Center of Nonlinear Science,\\ School of Mathematical and Physical Sciences, Wuhan Textile University, Wuhan 430200, China}
\author{Qin Zhou}
\email{qinzhou@whu.edu.cn}
\affiliation{Research Group of Nonlinear Optical Science and Technology, Research Center of Nonlinear Science,\\ School of Mathematical and Physical Sciences, Wuhan Textile University, Wuhan 430200, China}
\affiliation{State Key Laboratory of New Textile Materials and Advanced Processing Technologies, Wuhan Textile University, Wuhan 430200, China}
\author{B. A. Malomed}
\affiliation{Department of Physical Electronics, School of Electrical Engineering,
	Faculty of Engineering, and the Center for Light-Matter University, Tel Aviv
	University, Tel Aviv, Israel}
\affiliation{Instituto de Alta Investigaci\'{o}n, Universidad de Tarapac\'{a},
Casilla 7D, Arica, Chile}

\begin{abstract}
In the framework of the model of a spatially non-uniform Bose-Einstein
condensate with helicoidal spin-orbit (SO) coupling, we find abnormal
Peregrine solitons (PSs) on top of flat and periodic backgrounds, with
ultra-high amplitudes. We explore the roles of the SO coupling strength and
helicity pitch in the creation of these anomalously tall PSs and find that
their amplitude, normalized to the background height, attains indefinitely
large values. The investigation of the modulation instability (MI) in the
same system demonstrates that these PSs exist in a range of relatively weak
MI, maintaining the feasibility of their experimental observation.
\end{abstract}

\maketitle

\textit{Introduction.} Rogue waves (RWs), first discovered as extreme events
in the ocean~\cite%
{Dysthe2008,book,Chabchoub0,Chabchoub,Chabchoub2,Dudley2019}, have been
widely studied, due to their unique properties and potential applications,
in nonlinear optics~\cite{Solli2007,Lecaplain2012,Wenrong0,Mihalache0},
plasmas~\cite{Bailung2011}, Bose-Einstein condensates (BECs) \cite%
{Bludov2009,Romero-Ros2022,Romero-Ros2024}, magnetics \cite{magnetic},
financial markets \cite{financial}, and various other settings \cite%
{Zhenya1,Zhenya2,He0,He1,He,Chow0,Chow1,Chen2015,Chow2,Wenrong}. A widely
recognized RW prototype is provided by the exact Peregrine-soliton (PS)
solution of the nonlinear Schr\"{o}dinger equation (NLSE)~\cite%
{Peregrine1983}, whose characteristic features are the threefold peak
amplitude and spatiotemporal localization on top of the background~field
\cite{Tikan2017}. Several landmark experiments have directly demonstrated
this remarkable phenomenon and its ramifications~\cite%
{Romero-Ros2024,Tikan2017,Kibler2010,Michel2020}.

Spin-orbit (SO) coupling in BECs have drawn much interest since its
experimental implementation~\cite{Lin2011,Galitski2013,Hamner2015}, as it
offers the realization of the SO-coupling phenomenology in the uniquely
clean form \cite{Zhai,Engels} and make it possible to create artificial
vector gauge potentials~\cite{Dalibard2011,Ruseckas2005}. Recently, models
of BECs with non-uniform SO coupling have been introduced, as they provide
high tunability of this effect, and enhance the role of the intrinsic
nonlinearity in the SO-coupled BECs~\cite%
{Struck2012,Zhang2013,Jimenez-Garcia2015,Luo2016,Hejazi2020,Kartashov2014}.
In this context, soliton dynamics in the BEC with non-uniform landscapes of
the SO coupling has been investigated~\cite%
{Kartashov2017,Sherman,Bin-Liu,Bin-Liu2}, where, in particular, the
helicoidal gauge potential may originate from the light propagation in a
helical waveguide array~\cite{Rechtsman2013}. The propagation of matter-wave
solitons in a BEC with a random SO coupling was addressed~too \cite%
{Kartashov2019}.

SO-coupled BECs are modeled by systems of two (or several) coupled
Gross-Pitaevskii equations (GPEs). In this connection, it is relevant to
stress that PSs exist in multi-component NLSE models, such as the famous
Manakov system, but, due to the energy transfer between different
components, the PS amplitude is no longer fixed, although it still does not
exceed the triple background height~\cite%
{Chen2015,Baronio2012,Baronio2013,Chen2015a,Chen2015}. Nevertheless, recent
studies have shown that, under the action of self-steepening effects, the
amplitude of fundamental PSs can exceed the threefold limit, reaching up to
fivefold the background height~\cite{Chen2018}. In particular, exceptional
PSs, which feature ultra-high peak amplitudes, have been reported too in the
vector derivative NLSEs including the self-steepening effect~\cite{Chen2020}.

In this work, we focus on the following questions: can the fundamental PS
with an ultra-high peak amplitude be excited in other ways, besides using
higher-order effects, such as self-steepening, and to what extent is it
possible to increase the PS amplitude? To answer these questions, we first
consider a BEC model with non-uniform helicoidal SO coupling~(cf. Ref. \cite%
{Kartashov2017}), which offers experimental feasibility. We construct its
exact fundamental PS solutions on top of flat, alias continuous-wave (CW),
and periodic backgrounds. Through the analysis of the PS amplitude, we find
that PS with ultra-high peak amplitude, reaching indefinitely large values
(as normalized to the background height), can be created with the help of
the helicoidal SO coupling.

To explore the PS dynamics under the action of spatially non-uniform gauge
potentials, we consider the GPE for the spinor wave function $\mathbf{\Psi }%
=(\Psi _{1},\Psi _{2})^{T}$ of an effectively one-dimensional two-component
BEC, including the helicoidal SO coupling. In the scaled form (with $M=\hbar
=1$, where $M$ is the atomic mass), the GPE is ~\cite%
{Kartashov2017,Li2019,Li2021}
\begin{equation}
i\frac{\partial \mathbf{\Psi }}{\partial t}=\frac{1}{2}Q^{2}(x)\mathbf{\Psi }%
-(\mathbf{\Psi }^{\dag }\mathbf{\Psi })\mathbf{\Psi },
\label{helicoidal SOC}
\end{equation}%
where the helicoidally molded SO coupling is represented by the generalized
momentum operator,%
\begin{equation}
Q(x)=-i\partial _{x}+\alpha \boldsymbol{\sigma }\cdot \boldsymbol{n}(x).
\label{Q}
\end{equation}%
Here $\alpha $ is the SO-coupling strength, which is tunable in the
experiment \cite{Zhang2013,Jimenez-Garcia2015,Luo2016}, $\boldsymbol{\sigma }%
=(\sigma _{x},\sigma _{y},\sigma _{z})$ is the vector of the Pauli matrices,
and the spatial modulation is represented by vector%
\begin{equation}
\boldsymbol{n}(x)=(\cos (2\kappa x),\sin (2\kappa x),0),  \label{n}
\end{equation}%
with $\kappa <0$ and $\kappa >0$ corresponding to the left- and right-handed
helicity, respectively~\cite{Rechtsman2013,Samsonov2004,Burt2004}. As usual,
it is assumed that the inter- and intra-species attractive interactions have
equal strengths. Special forms of Eq.~(\ref{helicoidal SOC}) include the
uniform Rashba-Dresselhaus SO coupling \cite{Dalibard2011} when $\kappa =0$,
and the canonical Manakov system~\cite{Baronio2012} when $\alpha =0$.

\textit{Fundamental PS solutions.} Eq.~(\ref{helicoidal SOC}) is made
gauge-equivalent to the integrable Manakov system,%
\begin{equation}
i\mathbf{u}_{t}+\frac{1}{2}\mathbf{u}_{xx}+(\mathbf{u}^{\dag }\mathbf{u})%
\mathbf{u}=0,~~~\mathbf{u}=(u_{1},u_{2})^{T},  \label{Manakov system}
\end{equation}%
by means of the transformation \cite{Kartashov2019}
\begin{equation}
\mathbf{\Psi }=%
\begin{pmatrix}
\nu _{+}e^{-i(k_{\text{m}}+\kappa )x} & \nu _{-}e^{i(k_{\text{m}}-\kappa )x}
\\
\nu _{-}e^{-i(k_{\text{m}}-\kappa )x} & -\nu _{+}e^{i(k_{\text{m}}+\kappa )x}%
\end{pmatrix}%
\mathbf{u},  \label{trans1}
\end{equation}%
where $k_{\text{m}}=\sqrt{\alpha ^{2}+\kappa ^{2}}$ is the effective
momentum of the lowest-energy states, and
\begin{eqnarray}  \label{nu}
\nu _{+}=&\text{sgn}(\alpha )\sqrt{\left( k_{\text{m}}- \kappa \right)
/\left( 2k_{\text{m}}\right) } \\
\nu _{-}=&\sqrt{\left( k_{\text{m}}+ \kappa \right) /\left( 2k_{\text{m}%
}\right) }
\end{eqnarray}
Below, $k_{\text{m}}$ plays a crucial role determining properties of PSs,
especially as concerns the amplification of their amplitudes.

The Manakov system~(\ref{Manakov system}) possesses the Lax pair \cite%
{SVManakov} and admits the solution by means of the Darboux dressing method~%
\cite{Chen2015}. To begin with, we take the CW seed solution of Manakov
system~(\ref{Manakov system}), with components
\begin{equation}
u_{j0}=a\exp [-i(k_{j}x-\omega _{j}t)],~~j=1,2,  \label{seed}
\end{equation}%
which is determined by the amplitude ($a$), wavenumbers ($k_{j}$), and
frequencies
\begin{equation}
\omega _{j}=2a^{2}-k_{j}^{2}/2.  \label{dispersion relation}
\end{equation}%
Making use of the Manakov system invariance with respect to the rotation of
the set of the two components, we choose them in Eq. (\ref{seed}) with equal
amplitudes $a$. Subsequent results demonstrate that the helicoidal SO
coupling makes PS\ heights different in the two components $\Psi _{1,2}$ for
the same background amplitudes $a$, see Eqs. (\ref{background-periodic}) and
(\ref{background-cw}) below.

Utilizing the known PS solutions for Manakov system~(\ref{Manakov system})
derived by means of the Darboux transform~\cite{Chen2015}, and substitution~(%
\ref{trans1}), we obtain the following exact fundamental PS solutions of the
underlying Eq.~(\ref{helicoidal SOC}):
\begin{eqnarray}
\Psi _{1} &=&ae^{-i\kappa x}\left[ \nu _{+}\left( 1-\frac{\mathcal{R}_{1}}{%
\mathcal{N}_{1}}\right) e^{i\theta _{1}}+\nu _{-}\left( 1-\frac{\mathcal{R}%
_{2}}{\mathcal{N}_{2}}\right) e^{i\theta _{2}}\right] ,  \notag \\
\Psi _{2} &=&ae^{i\kappa x}\left[ \nu _{-}\left( 1-\frac{\mathcal{R}_{1}}{%
\mathcal{N}_{1}}\right) e^{i\theta _{1}}-\nu _{+}\left( 1-\frac{\mathcal{R}%
_{2}}{\mathcal{N}_{2}}\right) e^{i\theta _{2}}\right] ,  \notag \\
\theta _{1} &=&-(k_{\text{m}}+k_{1})x+\omega _{1}t,~~\theta _{2}=(k_{\text{m}%
}-k_{2})x+\omega _{2}t,  \label{rogue waves}
\end{eqnarray}%
where we define
\begin{eqnarray}
\mathcal{N}_{j} &=&\left[ (\theta +\mu t)^{2}+\zeta ^{2}t^{2}+\frac{4}{\zeta
^{2}}\right] \{[\delta +(-1)^{j}\mu ]^{2}+\zeta ^{2}\},  \notag \\
\mathcal{R}_{j} &=&8i\{\zeta ^{2}t-[\mu +(-1)^{j}\delta ](\theta +\mu
t)\}+16,  \notag \\
\mu &=&\pm \frac{\sqrt{2}}{2}\left[ \sqrt{\delta ^{2}(8a^2+\delta ^{2})}%
-4a^2+\delta ^{2}\right] ^{1/2},  \label{piecewise1}
\end{eqnarray}%
in the case of $|\delta |\geq a$, with $\delta \equiv k_{1}-k_{2}$, or
\begin{eqnarray}
\mathcal{N}_{j} &=&\left[ \theta ^{2}+(\zeta +\mu ^{\prime})^2t^{2}+\frac{4}{%
(\zeta +\mu ^{\prime})^2}\right] (2a^2+\zeta \mu ^{\prime}),  \notag \\
\mathcal{R}_{1} &=&4i(4a^2-\delta ^{2}+2\zeta \mu ^{\prime})t-4i(-1)^j\delta
\theta +8,  \notag \\
\mu ^{\prime } &=&\pm \frac{1}{\sqrt{2}}\left[ 4a^2-\delta ^{2}-\sqrt{\delta
^{2}(8a^2+\delta ^{2})}\right] ^{1/2},  \label{piecewise2}
\end{eqnarray}%
in the case of $|\delta |<a$. In either case, we set $\theta \equiv
2x+(k_{1}+k_{2})t$ and $\zeta \equiv \left( 1/\sqrt{2}\right) \left[ \sqrt{%
\delta ^{2}(8a^2+\delta ^{2})}+4a^2-\delta ^{2}\right] ^{1/2}$. Using the
translational symmetry, we shift the above solutions to the origin, to
produce compact expressions for them. Note that these PS solutions are
non-singular ones in the entire parameter range.

In addition to the same features which are demonstrated by the conventional
PSs, that exist in some multi-component systems, such as PSs of the
bright-dark type, PS doublets, \textit{etc}., the helicoidal SO coupling can
generate more intricate PS structures, among which the most salient aspect
is, as shown below, the possibility of having PSs with uniquely large heights.

The consideration of the exact solution~(\ref{rogue waves}) reveals that the
PS is generally located on top of a periodic background formed by the
superposition of two different CWs. The exact solution for the periodic
background is
\begin{eqnarray}
|\Psi _{1}^{\text{bg}}| &=&a\sqrt{1+\frac{\alpha }{k_{\text{m}}}\cos \left[
(\delta +2k_{\text{m}})x+\frac{k_{1}^{2}-k_{2}^{2}}{2}t\right] },  \notag \\
|\Psi _{2}^{\text{bg}}| &=&a\sqrt{1-\frac{\alpha }{k_{\text{m}}}\cos \left[
(\delta +2k_{\text{m}})x+\frac{k_{1}^{2}-k_{2}^{2}}{2}t\right] }.
\label{background-periodic}
\end{eqnarray}%
It is moving with speed $v=\left( k_{2}^{2}-k_{1}^{2}\right) /\left[
2(\delta +2k_{\text{m}})\right] $, where $k_{1,2}$ are the same wavenumbers
as in Eq. (\ref{seed}).

Note that, if wavenumbers $k_{1,2}$ and the momentum minimum $k_{\text{m}}$
satisfy the following relationship,
\begin{equation}
k_{1}=-k_{2}=-k_{\text{m}},  \label{relationship1}
\end{equation}%
the $\cos $ terms vanish in Eq.~(\ref{background-periodic}), i.e., the
periodic background degenerates into a flat CW. Due to the presence of the
helicoidal SO coupling, the constraint~(\ref{relationship1}) is different
from similar ones which provide for the flat background in the
coupled-NLSE~system \cite{Chen2015} and the multi-component
long-wave-short-wave resonance~model \cite{Chen2014aa}.

\textit{PS on the CW background.} \ To reveal the amplification effect of
the helicoidal SO coupling on the PS amplitude, we first address the PS
solution on top of the flat CW background, subject to constraint (\ref%
{relationship1}). The respective background amplitude~(\ref%
{background-periodic}) amounts to
\begin{equation}
|\Psi _{1}^{\text{cw}}|=a\sqrt{\frac{k_{\text{m}}+\alpha }{k_{\text{m}}}}%
,~|\Psi _{2}^{\text{cw}}|=a\sqrt{\frac{k_{\text{m}}-\alpha }{k_{\text{m}}}}.
\label{background-cw}
\end{equation}%
Under the action of the SO coupling with strength $\alpha $, the components
of the CW background (\ref{background-cw}) have different heights.

Taking into regard that the center of the PS solution (\ref{rogue waves}) is
pinned to the origin, enhancement factor $|F_{j}|$ of component $\Psi _{j}$
is defined as the peak-to-background ratio:
\begin{eqnarray}
F_{1} &=&\frac{\Psi _{1}(0,0)}{|\Psi _{1}^{\text{cw}}|}=\frac{a(\nu
_{+}f_{u_{1}}+\nu _{-}f_{u_{2}})}{|\Psi _{1}^{\text{cw}}|},  \notag \\
F_{2} &=&\frac{\Psi _{2}(0,0)}{|\Psi _{2}^{\text{cw}}|}=\frac{a(\nu
_{-}f_{u_{1}}-\nu _{+}f_{u_{2}})}{|\Psi _{2}^{\text{cw}}|},
\label{enhancement factors}
\end{eqnarray}%
where $|\Psi _{1,2}^{\text{cw}}|$ are given in Eq.~(\ref{background-cw}),
coefficients $\nu _{\pm }$ are same as in Eq. (\ref{nu}), and factors $%
f_{u_1}$ and $f_{u_{2}}$ are defined, for $|\delta |\geq a$, as
\begin{equation}
f_{u_{1}}=1-\frac{4\zeta ^{2}}{\zeta ^{2}+(\delta -\mu )^{2}},~~f_{u_{2}}=1-%
\frac{4\zeta ^{2}}{\zeta ^{2}+(\delta +\mu )^{2}},  \label{ratios1}
\end{equation}%
and, for $|\delta |<a$, as
\begin{equation}
f_{u_{1}}=f_{u_{2}}=1-\frac{2(\zeta +\mu ^{\prime })^{2}}{2a^2+\zeta \mu
^{\prime }}.  \label{ratios2}
\end{equation}

\begin{figure}[th]\vspace{-4mm}
\includegraphics[scale=0.35]{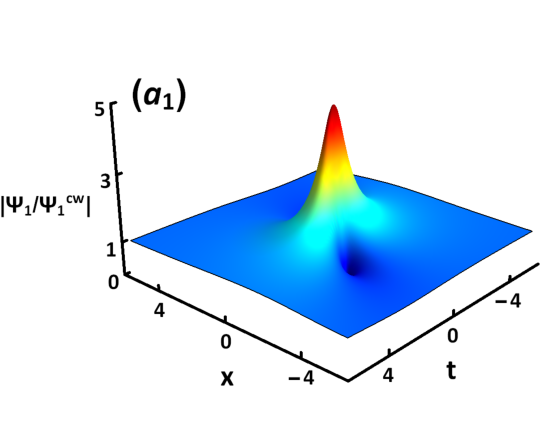}\hspace{3mm} %
\includegraphics[scale=0.35]{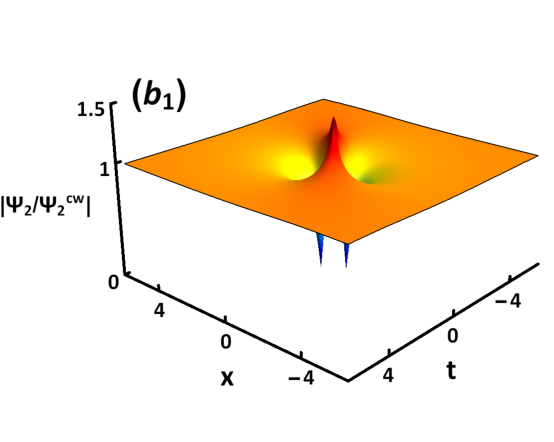} \newline
\vspace{-4mm} \includegraphics[scale=0.4]{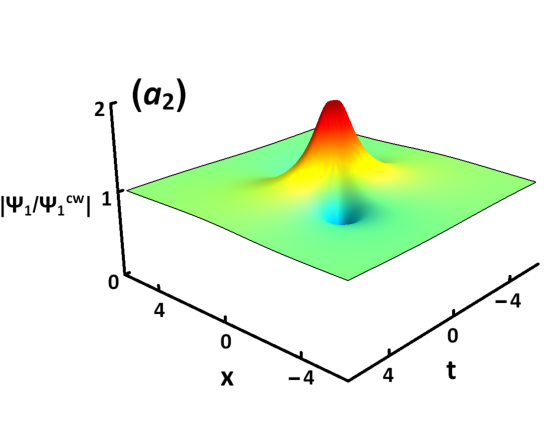}\hspace{3mm} %
\includegraphics[scale=0.35]{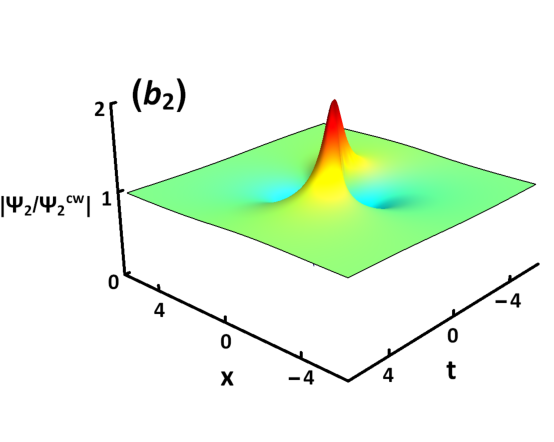} \newline
\vspace{-4mm} \includegraphics[scale=0.4]{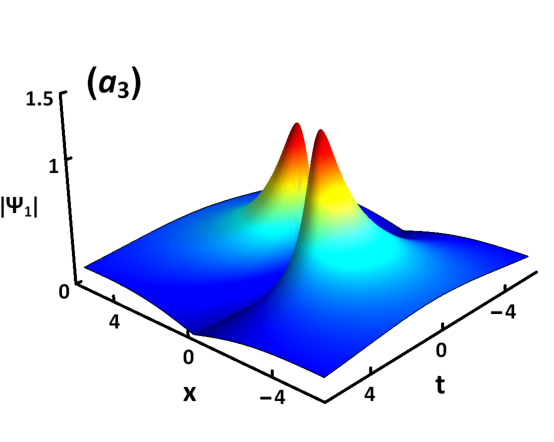}\hspace{3mm} %
\includegraphics[scale=0.35]{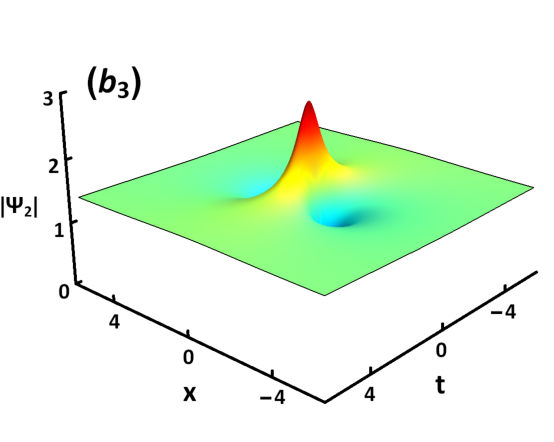}\newline
\vspace{-4mm}
\caption{$(a_{1},b_{1})$ An example of the fundamental PS , produced by
solution (\protect\ref{rogue waves}) under condition (\protect\ref%
{relationship1}), with the exceptionally high peak amplitude of the $\Psi
_{1} $ component, for $\protect\alpha =-1/2$, $\protect\kappa =2/5$. $%
(a_{2},b_{2})$ A generic PS in the Manakov system, for $\protect\alpha =0$. $%
(a_{3},b_{3})$ The PS with the zero background in $\Psi _{1}$ at $x=0$, for $%
\protect\alpha =-1/2$, $\protect\kappa =0$. The initial amplitude $a=1$.}
\label{fig1}
\end{figure}

Characteristic examples of the PSs featuring large enhancement factors are
presented in Fig.~\ref{fig1}, which includes a PS with nearly fivefold peak
amplitude for the component $\Psi _{1}$, with $\alpha =-1/2$ and $\kappa
=2/5 $, in Fig. \ref{fig1}($a_{1}$). For comparison, two special cases are
presented too, \textit{viz}., for $\alpha =0$ [Figs.~\ref{fig1}$(a_{2})$ and
$(b_{2})$] and $\kappa =0$ [Figs.~\ref{fig1}$(a_{3})$ and $(b_{3})$], which
correspond to the Manakov system limit and the uniform SO coupling,
respectively. It is observed that, with $\alpha =0$, the PS amplitudes are
only twice as large as those of the background (in fact, for the Manakov
system the peak amplitude cannot exceed three times the background~value
\cite{Chen2015}). In addition, for $\kappa =0$ the PS with zero background
in component $\Psi _{1}$ or $\Psi _{2} $ is produced by Eq.~(\ref%
{background-cw}), depending on the sign of $\alpha $. Thus, the helicoidal
SO coupling makes it possible to elevate the amplitude of one component to
an exceptional level, while suppressing the other component.

\begin{figure}[th]
\includegraphics[scale=0.65]{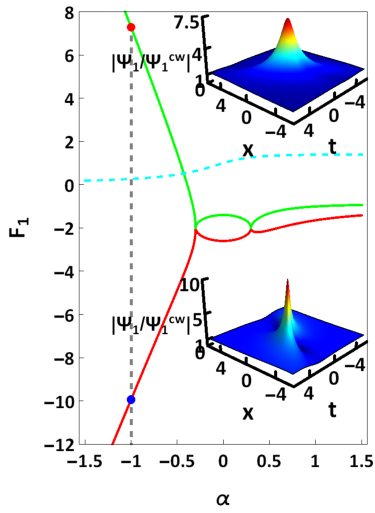}\hspace{1mm} %
\includegraphics[scale=0.65]{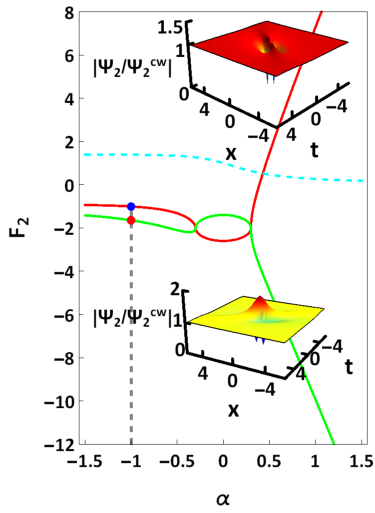}\newline
\vspace{-3mm} \vspace{-3mm}
\caption{Enhancement factors $F_{1}$ and $F_{2}$, as given by Eq. (\protect
\ref{enhancement factors}) for the PS with the flat background, vs. the
SO-coupling strength $\protect\alpha $ for $a=1$, $\protect\kappa %
=0.4,k_{1}=-k_{2}=-k_{\text{m}}$. The red and green curves correspond to $%
\protect\mu $ and $\protect\mu ^{\prime }$ taking signs $+$ or $-$ in Eqs. (%
\protect\ref{piecewise1}) and (\protect\ref{piecewise2}), respectively. The
cyan dashed curves show the CW background values $|\Psi _{j}^{\text{cw}}|$,
as given by Eq. (\protect\ref{background-cw}). The insets exhibit the
corresponding PSs at $\protect\alpha =-1$. }
\label{fig2}
\end{figure}

\begin{figure}[th]
\includegraphics[scale=0.65]{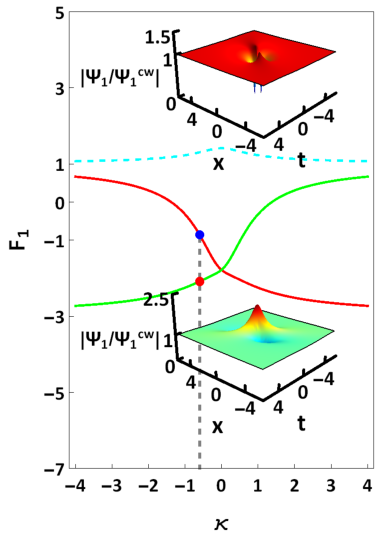}\hspace{1mm} %
\includegraphics[scale=0.65]{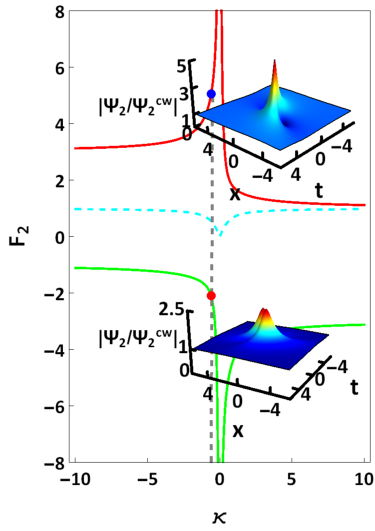}\newline
\vspace{-3mm} \vspace{-3mm}
\caption{Enhancement factors $F_{1}$ and $F_{2}$, as given by Eq. (\protect
\ref{enhancement factors}), vs. rotation frequency $\protect\kappa $ [see
Eq. (\protect\ref{n})] for $a=1$, $\protect\alpha =0.6,k_{1}=-k_{2}=-k_{%
\text{m}}$. The red and green curves correspond to $\protect\mu $ and $%
\protect\mu ^{\prime }$ taking signs $+$ or $-$ in Eqs. (\protect\ref%
{piecewise1}) and (\protect\ref{piecewise2}), respectively. The cyan dashed
curves show the CW background values $|\Psi _{j}^{\text{cw}}|$, as given by
Eq. (\protect\ref{background-cw}). The insets exhibit the corresponding PSs
at $\protect\kappa =-0.6$.}
\label{fig3}
\end{figure}

To further unveil the specific role of the helicoidal SO coupling in
generating PSs with exceptionally high amplitudes, we display the dependence
of enhancement factors $F_{j}$ on the SO-coupling strength $\alpha $ and
rotation frequency $\kappa $ in Figs.~\ref{fig2} and \ref{fig3},
respectively. They exhibit an indefinitely large (diverging) enhancement
factor for component $\Psi _{1}$ or $\Psi _{2}$ at $|\alpha |\rightarrow
\infty $ or $\kappa \rightarrow 0$. In particular, the insets to these
figures feature the enhancement factor $|F_{1}|$ with values close to $10$
at $\alpha =-1$ and $\kappa =0.4$, and $|F_{2}|$ close to $5$ at $\alpha =0.6
$ and $\kappa =-0.6$. A caveat is that the enhancement factor is diverging
when the background amplitude $|\Psi _{j}^{\text{CW}}|$ is vanishing, as
shown by the cyan dashed curves in Figs.~\ref{fig2} and \ref{fig3}. The
absolute values of the PS peak amplitude may be increased by taking values
of amplitude $a>1$ in Eq. (\ref{seed}) (recall it is currently fixed as $%
a\equiv 1$, by means of scaling).

\textit{PSs on top of the periodic background.} If the constraint (\ref%
{relationship1}) does not hold, the above solution~(\ref{rogue waves})
produces the PS built on top of the periodic background. Similar to the case
of the flat CW background considered above, we define the enhancement factor
to analyze the effect of the helicoidal SO coupling on the PSs. In Fig.~\ref%
{fig4} we demonstrate a characteristic example exceeding the threefold
contrast between the peak amplitude and periodic background in component $%
\Psi _{2}$, for $k_{1}=-k_{\text{m}},k_{2}=k_{\text{m}}-3/2$.

\begin{figure}[th]
\includegraphics[scale=0.4]{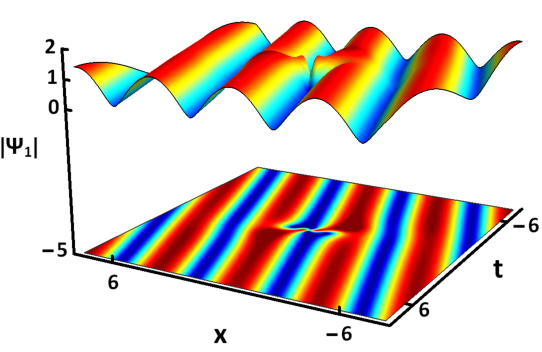}\hspace{3mm} %
\includegraphics[scale=0.4]{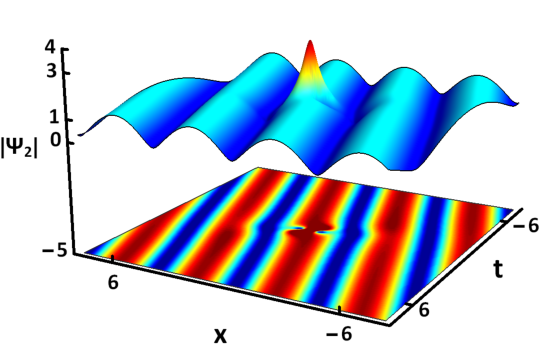}\newline
\vspace{-3mm} \vspace{-3mm}
\caption{The PS solution (\protect\ref{rogue waves}) built on top of the
periodic background. It exceeds the exceeding the threefold enhancement
limit, with $k_{1}=-k_{\text{m}}$ and $k_{2}=k_{\text{m}}-3/2$. The other
parameters are $a=1,\protect\alpha =-1,\protect\kappa =0.4$.}
\label{fig4}
\end{figure}

\begin{figure}[th]
\includegraphics[scale=0.35]{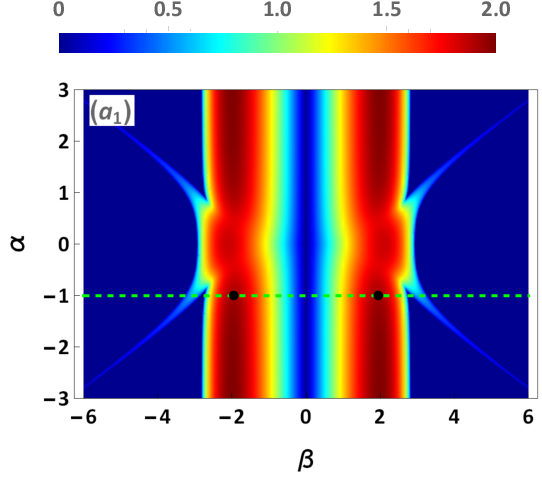}\hspace{3mm} %
\includegraphics[scale=0.35]{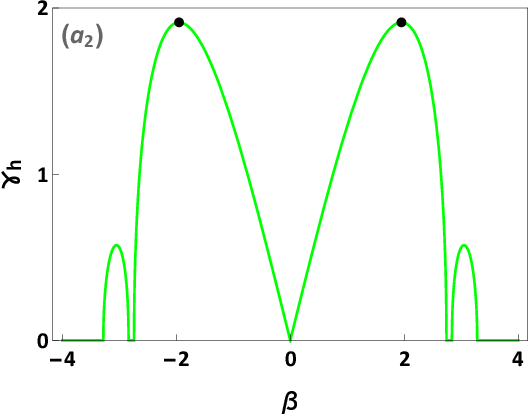} \newline
\includegraphics[scale=0.35]{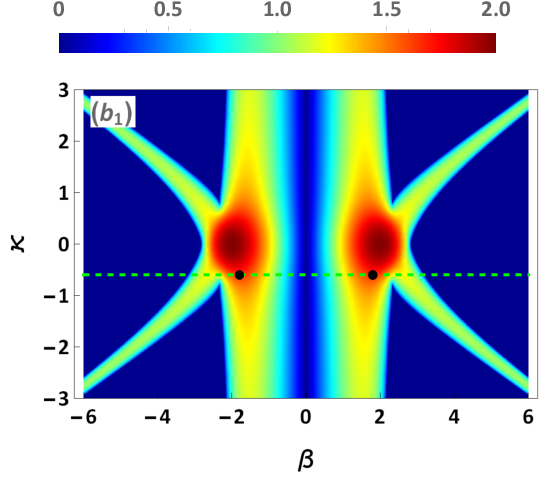}\hspace{3mm} %
\includegraphics[scale=0.35]{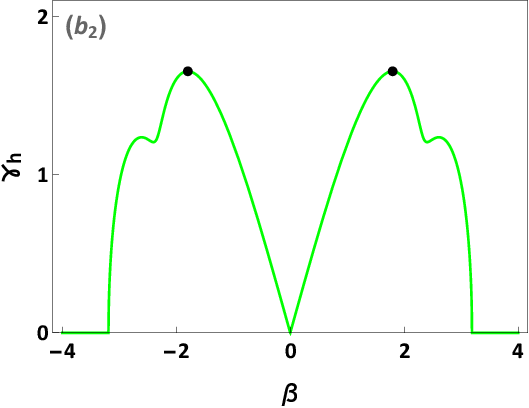}\newline
\vspace{-3mm} \vspace{-3mm}
\caption{Heatmaps of the MI gain $\protect\gamma _{\mathrm{h}}$ in the $(%
\protect\beta ,\protect\alpha )$ plane $(a_{1})$ for $\protect\kappa =0.4$,
and in the $(\protect\beta ,\protect\kappa )$ plane $(b_{1})$ for $\protect%
\alpha =0.6$. Panels $(a_{2})$ and $(b_{2})$ exhibit, respectively, the gain
profile $\protect\gamma _{\mathrm{h}}$ at $\protect\alpha =-1$ and $\protect%
\kappa =-0.6$, with the maximum gain marked by black dots. The amplitudes of
the underlying CW state are $a_{1}=a_{2}=1$.}
\label{fig5}
\end{figure}

Next, we address the modulation instability (MI) of the CW field, $\Psi
_{j0}=a_{j}e^{i\mu t}$ with $\mu =a_{1}^{2}+a_{2}^{2}$, where $a_{1}$ and $%
a_{2}$ are the uniform amplitudes and $\mu $ is the chemical potential, in
the presence of the helicoidal SO-coupled BECs. To this end, we add small
perturbations to the CW fields, \textit{viz}., $\Psi _{j}=\Psi
_{j0}\{1+p_{j}\exp [-i(\beta x-\Omega t)]+q_{j}^{\ast }\exp [i(\beta
x-\Omega ^{\ast }t)]\}$, where $\beta $ and $\Omega $ are, respectively, the
real and complex parameters, $p_{j}$ and $q_{j}$ being small complex
amplitudes. Linearizing the corresponding Eq.~(\ref{helicoidal SOC}) with
respect to $p_{j}$ and $q_{j}$, we derive a quartic equation for the
perturbation eigenfrequency $\Omega $, which determines the MI gain as $%
\gamma _{\mathrm{h}}={|\text{Im}(\Omega )|}_{\text{max}}$. In Fig.~\ref{fig5}%
, we display heatmaps for the so found value of $\gamma _{h}$ in the $(\beta
,\alpha )$ and $(\beta ,\kappa )$ parameter planes. The plots reveal that
the MI-gain spectra are symmetrically distributed in broad regions of $%
\alpha $ and $\kappa $, which can give rise to the anomalous PS behavior in
a broad range of parameters, in comparison to the usual situation underlain
by the baseband-MI analysis~\cite{Baronio2014,Baronio2015}. For instance,
Figs.~\ref{fig5}$(a_{2})$ and $\left( b_{2}\right) $ demonstrate,
respectively, that the gain maximum, $\gamma _{\mathrm{h}}\approx 1.91$ at $%
\alpha =-1$ and $\kappa =0.4$, corresponds to the enhancement factor $%
|F_{1}|\approx 10$ in Fig.~\ref{fig2}, and the maximum $\gamma _{\mathrm{h}%
}\approx 1.65$, at $\alpha =0.6$ and $\kappa =-0.6$, corresponds to $%
|F_{2}|\approx 5$ in Fig.~\ref{fig3}. Such relatively small values of the MI
gain, corresponding to the ultra-high PS peak amplitudes, suggest that these
large amplitude values may be relatively easy to attain in the experiment,
as the background will not be vulnerable to the quick destruction by of the
MI-driven blowup, hence these PSs are rather robust modes.

\begin{figure}[th]\vspace{-7mm}
\includegraphics[scale=0.28]{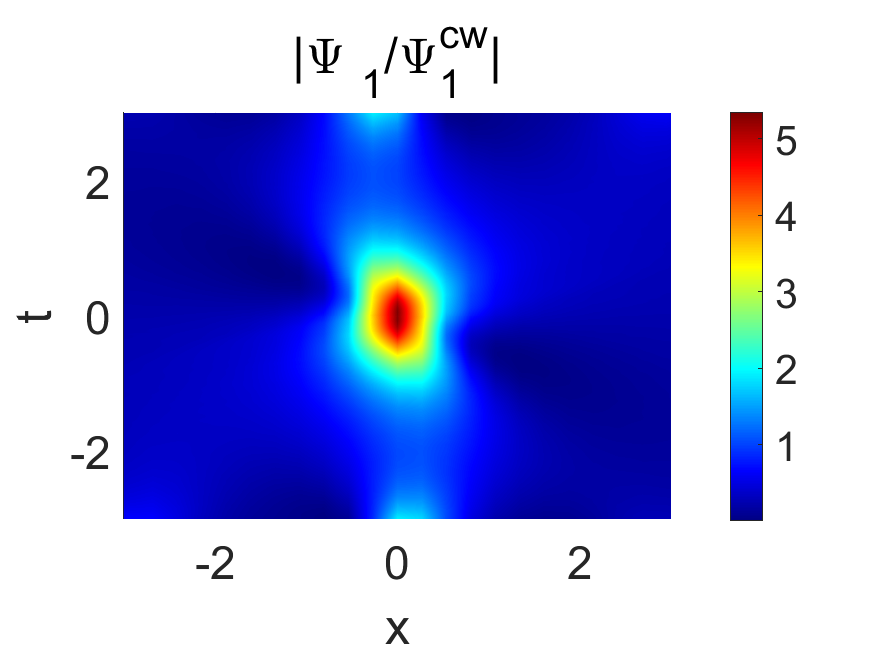}\hspace{1mm} %
\includegraphics[scale=0.28]{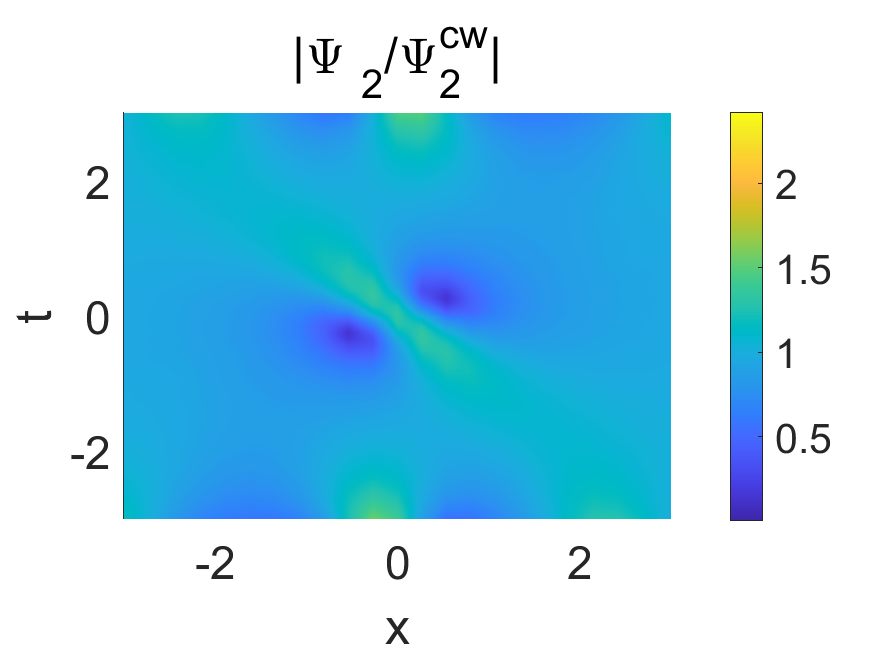}\newline
\vspace{-3mm} \vspace{-3mm}
\caption{The result of the numerical simulations of the fundamental PSs from
Figs. \protect\ref{fig1}($a_{1}$) and ($b_{1}$) under the action of the $2\%$
noise.}
\label{fig6}
\end{figure}

To test the expected robustness of the PSs in the present setting, in Fig.~%
\ref{fig6} we display results of the numerically simulated evolution of the
PSs from Figs. \ref{fig1}($a_{1}$) and ($b_{1}$) under the action of $2\%$
random disturbances. It is observed that the PSs with the ultra-high
amplitudes indeed demonstrate robust propagation.

\textit{Conclusion.} We have reported the occurrence of abnormal fundamental
PSs (Peregrine solitons) with ultra-high peak amplitudes in the integrable
system of GPEs (Gross-Pitaevskii equations) including the helicoidally
modulated SO (spin-orbit) coupling, which is a gauge isomer of the Manakov
system. The PS solutions are found on top of both the flat and periodic-wave
backgrounds. The results demonstrate the existence of RWs (rogue waves with
the ultra-high amplitude in the context of matter waves (BEC)), while
previously this was reported in models of nonlinear optics~\cite%
{Chen2018,Chen2020}. The helicoidal SO coupling is crucially important for
generating this abnormal PSs, and the controllable nature of the SO coupling
makes the predicted phenomenology experimentally feasible. The MI
(modulational instability) is also studied in the system, demonstrating that
the high-amplitude PSs readily coexist with moderate MI, thus preventing a
strong background instability and improving chances for the experimental
creation of the predicted tall rogue waves.


\textit{Acknowledgments} This work was supported by the National Natural Science
Foundation of China (Grant No. 11975172) and the Natural Science Foundation of Hubei Province of China (Grant No.~2023AFB222). The work of B.A.M. is
supported, in part, by the Israel Science Foundation (Grant No. 1695/22).


\begin{thebibliography}{0}%
\makeatletter
\providecommand \@ifxundefined [1]{%
 \@ifx{#1\undefined}
}%
\providecommand \@ifnum [1]{%
 \ifnum #1\expandafter \@firstoftwo
 \else \expandafter \@secondoftwo
 \fi
}%
\providecommand \@ifx [1]{%
 \ifx #1\expandafter \@firstoftwo
 \else \expandafter \@secondoftwo
 \fi
}%
\providecommand \natexlab [1]{#1}%
\providecommand \enquote  [1]{``#1''}%
\providecommand \bibnamefont  [1]{#1}%
\providecommand \bibfnamefont [1]{#1}%
\providecommand \citenamefont [1]{#1}%
\providecommand \href@noop [0]{\@secondoftwo}%
\providecommand \href [0]{\begingroup \@sanitize@url \@href}%
\providecommand \@href[1]{\@@startlink{#1}\@@href}%
\providecommand \@@href[1]{\endgroup#1\@@endlink}%
\providecommand \@sanitize@url [0]{\catcode `\\12\catcode `\$12\catcode
  `\&12\catcode `\#12\catcode `\^12\catcode `\_12\catcode `\%12\relax}%
\providecommand \@@startlink[1]{}%
\providecommand \@@endlink[0]{}%
\providecommand \url  [0]{\begingroup\@sanitize@url \@url }%
\providecommand \@url [1]{\endgroup\@href {#1}{\urlprefix }}%
\providecommand \urlprefix  [0]{URL }%
\providecommand \Eprint [0]{\href }%
\providecommand \doibase [0]{https://doi.org/}%
\providecommand \selectlanguage [0]{\@gobble}%
\providecommand \bibinfo  [0]{\@secondoftwo}%
\providecommand \bibfield  [0]{\@secondoftwo}%
\providecommand \translation [1]{[#1]}%
\providecommand \BibitemOpen [0]{}%
\providecommand \bibitemStop [0]{}%
\providecommand \bibitemNoStop [0]{.\EOS\space}%
\providecommand \EOS [0]{\spacefactor3000\relax}%
\providecommand \BibitemShut  [1]{\csname bibitem#1\endcsname}%
\let\auto@bib@innerbib\@empty
\end{thebibliography}%


\begin{thebibliography}{99}
\bibitem{Dysthe2008} K. Dysthe, H. E. Krogstad, P. M\"{u}ller, Oceanic rogue
waves, Annual Rev. Fluid Mech. \textbf{40}, 287 (2008).

\bibitem{book} C. Kharif, E. Pelinovsky, and A. Slunyaev, \textit{Rogue
Waves in the Ocean} (Springer-Verlag: Berlin, 2009).

\bibitem{Chabchoub0} A. Chabchoub, N. P. Hoffmann, and N. Akhmediev, Rogue
wave observation in a water wave tank, Phys. Rev. Lett. \textbf{106}, 204502
(2011).

\bibitem{Chabchoub} A. Chabchoub, N. Hoffmann, M. Onorato, A. Slunyaev, A.
Sergeeva, E. Pelinovsky, and N. Akhmediev, Observation of a hierarchy of up
to fifth-order rogue waves in a water tank, Phys. Rev. E \textbf{86}, 056601
(2012).

\bibitem{Chabchoub2} A. Chabchoub, N. Hoffmann, M. Onorato, and N.
Akhmediev, Super rogue waves: Observation of a higher-order breather in
water waves. Phys. Rev. X \textbf{2}, 011015 (2012).

\bibitem{Dudley2019} J. M. Dudley, G. Genty, A. Mussot, A. Chabchoub, F.
Dias, Rogue waves and analogies in optics and oceanography, Nature Rev.
Phys. \textbf{1}, 675 (2019).

\bibitem{Solli2007} D. R. Solli, C. Ropers, P. Koonath, B. Jalali, Optical
rogue waves, nature, \textbf{450}, 1054 (2007).

\bibitem{Lecaplain2012} C. Lecaplain, P. Grelu, J. M. Soto-Crespo, N.
Akhmediev, Dissipative rogue waves generated by chaotic pulse bunching in a
mode-locked laser, Phys. Rev. Lett. \textbf{108}, 233901 (2012).

\bibitem{Wenrong0} W. R. Sun, B. Tian, Y. Jiang, and H. L. Zhen, Optical
rogue waves associated with the negative coherent coupling in an isotropic
medium, Phys. Rev. E \textbf{91}, 023205 (2015).

\bibitem{Mihalache0} S. H. Chen, J. M. Soto-Crespo, F. Baronio, P. Grelu,
and D. Mihalache, Rogue-wave bullets in a composite (2+1)D nonlinear medium,
Opt. Exp. \textbf{24}, 15251-15260 (2016).

\bibitem{Bailung2011} H. Bailung, S. K. Sharma, Y. Nakamura, Observation of
Peregrine solitons in a multicomponent plasma with negative ions, Phys. Rev.
Lett. \textbf{107}, 255005 (2011).

\bibitem{Bludov2009} Y. V. Bludov, V. V. Konotop, N. Akhmediev, Matter rogue
waves, Phys. Rev. A \textbf{80}, 033610 (2009).

\bibitem{Romero-Ros2022} A. Romero-Ros, G. C. Katsimiga, S. I. Mistakidis,
B. Prinari, G. Biondini, P. Schmelcher, P. G. Kevrekidis, Theoretical and
numerical evidence for the potential realization of the Peregrine soliton in
repulsive two-component Bose-Einstein condensates, Phys. Rev. A \textbf{105}%
, 053306 (2022).

\bibitem{Romero-Ros2024} A. Romero-Ros, G. C. Katsimiga, S. I. Mistakidis,
S. Mossman, G. Biondini, P. Schmelcher, P. Engels, P. G. Kevrekidis,
Experimental realization of the Peregrine soliton in repulsive two-component
Bose-Einstein condensates, Phys. Rev. Lett. \textbf{132}, 033402 (2024).

\bibitem{magnetic} W. R. Sun, B. Tian, H. L. Zhen, and Y. Sun, Breathers and
rogue waves of the fifth-order nonlinear Schr\"{o}dinger equation in the
Heisenberg ferromagnetic spin chain, Nonlin. Dynamics \textbf{81}, 725-732
(2015).

\bibitem{financial} Z. Y. Yan, Financial rogue waves, Comm. Theor. Phys.
\textbf{54}, 947-949 (2010).

\bibitem{Zhenya1} Z. Y. Yan, Nonautonomous \textquotedblleft rogons" in the
inhomogeneous nonlinear Schr\"{o}dinger equation with variable coefficients,
Phys. Lett. A \textbf{374}, 672-679 (2010).

\bibitem{Zhenya2} Z. Yan, V. V. Konotop, and N. Akhmediev, Three-dimensional
rogue waves in nonstationary parabolic potentials, Phys. Rev. E \textbf{82},
036610 (2010).

\bibitem{He0} S. W. Xu, J. S. He, and L. H. Wang, The Darboux transformation
of the derivative nonlinear Schr\"{o}dinger equation, J. Phys. A: Math. Gen.
\textbf{44}, 305203 (2011).

\bibitem{He1} Y. S. Tao and J. S. He, Multisolitons, breathers, and rogue
waves for the Hirota equation generated by the Darboux transformation, Phys.
Rev. E \textbf{85}, 026601 (2012).

\bibitem{He} J. S. He, H. R. Zhang, L. H. Wang, K. Porsezian, and A. S.
Fokas, Generating mechanism for higher-order rogue waves, Phys. Rev. E
\textbf{87}, 052914 (2013).

\bibitem{Chow0} K. W. Chow, H. N. Chan, D. J. Kedziora, and R. H. J.
Grimshaw, Rogue wave modes for the long wave-short wave resonance model, J.
Phys. Soc. Jpn. \textbf{82}, 074001 (2013).

\bibitem{Chow1} H. N. Chan, K. W. Chow, D. J. Kedziora, R. H. J. Grimshaw,
and E. Ding, Rogue wave modes for a derivative nonlinear Schr\"{o}dinger
model, Phys. Rev. E \textbf{89}, 032914 (2014).

\bibitem{Chen2015} S. Chen, D. Mihalache, Vector rogue waves in the Manakov
system: diversity and compossibility, J. Phys. A \textbf{48}, 215202 (2015).

\bibitem{Chow2} J. G. Rao, K. W. Chow, D. Mihalache, and J. S. He,
Completely resonant collision of lumps and line solitons in the
Kadomtsev-Petviashvili I equation, Stud. Appl. Math. \textbf{127}, 1007-1035
(2021).

\bibitem{Wenrong} L. Liu, W.-R. Sun, and B. A. Malomed, Formation of rogue
waves and modulational instability with zero-wavenumber gain in
multi-component systems with coherent coupling, Phys. Rev. Lett. \textbf{131}%
, 093801 (2023).

\bibitem{Peregrine1983} D. H. Peregrine, Water waves, nonlinear Schr\"{o}%
dinger equations and their solutions, J. Aust. Math. Soc. Ser. B \textbf{25}%
, 16 (1983).

\bibitem{Tikan2017} A. Tikan, C. Billet, G. El, A. Tovbis, M. Bertola, T.
Sylvestre, F. Gustave, S. Randoux, G. Genty, P. Suret, J. M. Dudley,
Universality of the Peregrine soliton in the focusing dynamics of the cubic
nonlinear Schr\"{o}dinger equation, Phys. Rev. Lett. \textbf{119}, 033901
(2017).

\bibitem{Kibler2010} B. Kibler, J. Fatome, C. Finot, G. Millot, F. Dias, G.
Genty, N. Akhmediev, J. M. Dudley, The Peregrine soliton in nonlinear fibre
optics, Nature Phys. \textbf{6}, 790 (2010).

\bibitem{Michel2020} G. Michel, F. Bonnefoy, G. Ducrozet, G. Prabhudesai, A.
Cazaubiel, F. Copie, A. Tikan, P. Suret, S. Randoux, E. Falcon, Emergence of
Peregrine solitons in integrable turbulence of deep water gravity waves,
Phys. Rev. Fluids \textbf{5}, 082801 (2020).

\bibitem{Lin2011} Y. J. Lin, K. Jim\'{e}nez-Garc\'{\i}a, I. B. Spielman,
Spin-orbit-coupled Bose-Einstein condensates, Nature (London) \textbf{471},
83 (2011).

\bibitem{Galitski2013} V. Galitski, I. B. Spielman, Spin-orbit coupling in
quantum gases, Nature (London) \textbf{494}, 49 (2013).

\bibitem{Hamner2015} C. Hamner, Y. Zhang, M. A. Khamehchi, M. J. Davis, P.
Engels, Spin-Orbit-Coupled Bose-Einstein Condensates in a One-Dimensional
Optical Lattice, Phys. Rev. Lett. \textbf{114}, 070401 (2015).

\bibitem{Zhai} H. Zhai, Degenerate quantum gases with spin--orbit coupling:
a review, Rep. Prog. Phys. \textbf{78}, 026001 (2015).

\bibitem{Engels} Y. Zhang, M. E. Mossman, T. Busch, P. Engels, and C. Zhang,
Properties of spin--orbit-coupled Bose--Einstein condensates, Front. Phys.
\textbf{11}, 118103 (2016).

\bibitem{Dalibard2011} J. Dalibard, F. Gerbier, G. Juzeli\={u}nas, P. \"{O}%
hberg, Colloquium: Artificial gauge potentials for neutral atoms, Rev. Mod.
Phys. \textbf{83}, 1523 (2011).

\bibitem{Ruseckas2005} J. Ruseckas, G. Juzeli\={u}nas, P. \"{O}hberg, and M.
Fleischhauer, Non-Abelian Gauge Potentials for Ultracold Atoms with
Degenerate Dark States, Phys. Rev. Lett. \textbf{95}, 010404 (2005).

\bibitem{Struck2012} J. Struck, C. \"{O}lschl\"{a}ger, M. Weinberg, P.
Hauke, J. Simonet, A. Eckardt, M. Lewenstein, K. Sengstock, P.
Windpassinger, Tunable Gauge Potential for Neutral and Spinless Particles in
Driven Optical Lattices, Phys. Rev. Lett. \textbf{108}, 225304 (2012).

\bibitem{Zhang2013} Y. Zhang, G. Chen, and C. Zhang, Tunable spin-orbit
coupling and quantum phase transition in a trapped Bose-Einstein condensate,
Sci. Rep. \textbf{3}, 1937 (2013).

\bibitem{Jimenez-Garcia2015} K. Jim\'{e}nez-Garc\'{\i}a, L. J. LeBlanc, R.
A. Williams, M. C. Beeler, C. Qu, M. Gong, C. Zhang, I. B. Spielman, Tunable
Spin-Orbit Coupling via Strong Driving in Ultracold-Atom Systems, Phys. Rev.
Lett. \textbf{114}, 125301 (2015).

\bibitem{Luo2016} X. Luo, L. Wu, J. Chen, Q. Guan, K. Gao, Z.-F. Xu, L. You,
R. Wang, Tunable atomic spin-orbit coupling synthesized with a modulating
gradient magnetic field, Sci. Rep. \textbf{6}, 18983 (2016).

\bibitem{Hejazi2020} S. S. S. Hejazi, J. Polo, R. Sachdeva, T. Busch,
Symmetry breaking in binary Bose-Einstein condensates in the presence of an
inhomogeneous artificial gauge field, Phys. Rev. A \textbf{102}, 053309
(2020).

\bibitem{Kartashov2014} Y. V. Kartashov, V. V. Konotop, and D. A. Zezyulin,
Bose-Einstein condensates with localized spin-orbit coupling: Soliton
complexes and spinor dynamics, Phys. Rev. A \textbf{90}, 063621 (2014).

\bibitem{Kartashov2017} Y. V. Kartashov and V. V. Konotop, Solitons in
Bose-Einstein Condensates with Helicoidal Spin-Orbit Coupling, Phys. Rev.
Lett. \textbf{118}, 190401 (2017).

\bibitem{Sherman} Y. V. Kartashov, E. Sherman, B. Malomed, and V. Konotop,
Stable two-dimensional soliton complexes in Bose-Einstein condensates with
helicoidal spin-orbit coupling, New J. Phys. 22, 103014 (2020).

\bibitem{Bin-Liu} Y. Li, X. Zhang, R. Zhong, Z. Luo, B. Liu, C. Huang, W.
Pang, B. A. Malomed, Two-dimensional composite solitons in Bose-Einstein
condensates with spatially confined spin-orbit coupling, Comm. Nonlin. Sci.
Num. Sim. \textbf{73}, 481-489 (2019).

\bibitem{Bin-Liu2} B. Liu, R. Zhong, Z. Chen, X. Qin, H. Zhong, Y. Li and B.
A. Malomed, Holding and transferring matter-wave solitons against gravity by
spin-orbit-coupling tweezers, New J. Phys. \textbf{22}, 043004 (2020).

\bibitem{Rechtsman2013} M. C. Rechtsman, J. M. Zeuner, Y. Plotnik, Y. Lumer,
D. Podolsky, F. Dreisow, S. Nolte, M. Segev, A. Szameit, Photonic Floquet
topological insulators, Nature (London) \textbf{496}, 196 (2013).

\bibitem{Kartashov2019} Y. V. Kartashov, V. V. Konotop, M. Modugno, E. Ya.
Sherman, Solitons in Inhomogeneous Gauge Potentials: Integrable and
Nonintegrable Dynamics, Phys. Rev. Lett. \textbf{122}, 064101 (2019).

\bibitem{Baronio2012} F. Baronio, A. Degasperis, M. Conforti, S. Wabnitz,
Solutions of the vector nonlinear Schr\"{o}dinger equations: evidence for
deterministic rogue waves, Phys. Rev. Lett. \textbf{109}, 044102 (2012).

\bibitem{SVManakov} S. V. Manakov, On the theory of two-dimensional
stationary self-focusing of electromagnetic waves, Zh. Eksp. Teor. Fiz.
\textbf{65}, 505-516 (1973) [Sov. Phys. JETP \textbf{38}, 248-253 (1974)].

\bibitem{Baronio2013} F. Baronio, M. Conforti, A. Degasperis, S. Lombardo,
Rogue waves emerging from the resonant interaction of three waves, Phys.
Rev. Lett. \textbf{111}, 114101 (2013).

\bibitem{Chen2015a} S. Chen, F. Baronio, J. M. Soto-Crespo, Ph. Grelu, M.
Conforti, S. Wabnitz, Optical rogue waves in parametric three-wave mixing
and coherent stimulated scattering, Phys. Rev. A \textbf{92}, 033847 (2015).

\bibitem{Chen2018} S. Chen, Y. Ye, J. M. Soto-Crespo, P. Grelu, F. Baronio,
Peregrine Solitons Beyond the Threefold Limit and Their Two-Soliton
Interactions, Phys. Rev. Lett. \textbf{121}, 104101 (2018).

\bibitem{Chen2020} S. Chen, C. Pan, P. Grelu, F, Baronio, N. Akhmediev,
Fundamental Peregrine Solitons of Ultrastrong Amplitude Enhancement through
Self-Steepening in Vector Nonlinear Systems, Phys. Rev. Lett. \textbf{124},
113901 (2020).

\bibitem{Li2019} X. X. Li, R. J. Cheng, A. X. Zhang, J. K. Xue, Modulational
instability of Bose-Einstein condensates with helicoidal spin-orbit
coupling, Phys. Rev. E \textbf{100}, 032220 (2019).

\bibitem{Li2021} X. X. Li, R. J. Cheng, J. L. Ma, A. X. Zhang, J. K. Xue,
Solitary matter wave in spin-orbit-coupled Bose-Einstein condensates with
helicoidal gauge potential, Phys. Rev. E \textbf{104}, 034214 (2021).

\bibitem{Samsonov2004} S. V. Samsonov, A. D. R. Phelps, V. L. Bratman, G.
Burt, G. G. Denisov, A. W. Cross, K. Ronald, W. He, H. Yin, Compression of
frequency-modulated pulses using helically corrugated waveguides and its
potential for generating multigigawatt rf radiation, Phys. Rev. Lett.
\textbf{92}, 118301 (2004).

\bibitem{Burt2004} G. Burt, S. V. Samsonov, K. Ronald, G. G. Denisov, A. R.
Young, V. L. Bratman, A. D. R. Phelps, A. W. Cross, I. V. Konoplev, W. He,
J. Thomson, C. G. Whyte, Dispersion of helically corrugated waveguides:
Analytical, numerical, and experimental study, Phys. Rev. E \textbf{70},
046402 (2004).

\bibitem{Chen2014aa} S. Chen, J. M. Soto-Crespo, P. Grelu, Coexisting rogue
waves within the (2+1)-component long-wave--short-wave resonance, Phys. Rev.
E \textbf{90}, 033203 (2014).

\bibitem{Baronio2014} F. Baronio, M. Conforti, A. Degasperis, S. Lombardo,
M. Onorato, S. Wabnitz, Vector Rogue Waves and Baseband Modulation
Instability in the Defocusing Regime, Phys. Rev. Lett. \textbf{113}, 034101
(2014).

\bibitem{Baronio2015} F. Baronio, S. Chen, Ph. Grelu, S. Wabnitz, M.
Conforti, Baseband modulation instability as the origin of rogue waves,
Phys. Rev. A \textbf{91}, 033804 (2015).
\end{thebibliography}

\end{document}